\documentclass[useAMS,usenatbib,twocolumn]{mn2e}
\usepackage{graphicx}
\usepackage{natbib}

\title[Deuterium enrichment of the interstellar grain mantle]
{Deuterium enrichment of the interstellar grain mantle}
\author[Ankan Das, Dipen Sahu, Liton Majumdar, Sandip K. Chakrabarti]
{Ankan Das$^{1}$, Dipen Sahu$^{1}$, Liton Majumdar$^{2,3,1}$, Sandip K. Chakrabarti$^{4,1}$\\
$^{1}$Indian Centre for Space Physics, Chalantika 43, Garia Station Rd.,
             Kolkata, 700084, India\\
$^{2}$ Univ. Bordeaux, LAB, UMR 5804, F-33270, Floirac, France\\
$^3$ CNRS, LAB, UMR 5804, F-33270, Floirac, France\\
$^{4}$S. N. Bose National Centre for Basic Sciences, Salt Lake,
              Kolkata 700098, India}
\begin{document}

\date{}


\maketitle


\begin{abstract}
We carry out Monte-Carlo simulation to study deuterium enrichments of interstellar
grain mantles under various physical conditions. Based on the physical properties, various
types of clouds are considered. We find that in diffuse cloud regions,
very strong radiation fields persists and hardly a few layers of surface species are
formed. In translucent cloud regions with a moderate radiation field, significant number of layers would be
produced and surface coverage is mainly dominated by photo-dissociation products such as,
$\mathrm{C,CH_3,CH_2D,OH}$ and OD. In the intermediate dense cloud regions
(having number density of total hydrogen nuclei
in all forms $\sim 2 \times 10^4$ cm$^{-3}$), water and methanol along with their deuterated
derivatives are efficiently formed. For much higher density regions
($\sim 10^6$ cm$^{-3}$), water and methanol productions are suppressed but surface
coverages of $\mathrm{CO,CO_2,O_2,O_3}$ are dramatically increased.
We find a very high degree of fractionation of water and methanol.
Observational results support a high fractionation of methanol but surprisingly
water fractionation is found to be low. This is in contradiction with our model
results indicating alternative routes for de-fractionation of water. 
Effects of various types of energy barriers are also studied. 
Moreover, we allow grain mantles to interact with various charged particles
(such as $\mathrm{H^+,\ Fe^+,S^+}$ and C$^+$) to study the stopping power and projected
range of these charged particles on various target ices.
\end{abstract}


\section{Introduction}

Astronomical observations along with laboratory studies reveal the presence of 
numerous organic molecules in interstellar space \citep{herb09,orge04,abel66}. 
Modeling of interstellar composition also supports indigenous formation of these molecules
\citep{das08a,chak15,maju12,maju13,das15a}.
Modeling complements observational and laboratory studies by emphasizing
the role of interstellar dusts towards the formation of these molecules. 
Mainly silicate and carbonaceous grains dominate in an interstellar space and they are thought to 
have molecular ice layers \citep{drai03,gibb04}. 
In fact, even the huge abundance of molecular hydrogen could not be 
explained without invoking interstellar dusts \citep{biha01,chan05,chak06a,chak06b,sahu15}.
It is now believed that at least some significant fraction of the
interstellar species would be synthesized on interstellar dusts \citep{tiel82,das10,das11,das08b}.

Stars are mainly formed in the densest part of an interstellar cloud called as 
pre-stellar core. These regions are extremely cold ($< 10$K) and dense ($> 10^5$ cm$^{-3}$). 
At these temperatures and densities, gas phase species could stick to a grain surface and 
lighter species could migrate through the surface either by thermal hopping or by 
quantum mechanical tunnelling. Earlier studies for the formations of 
interstellar grain mantle by various authors 
\citep{stan04,das10,das11} pointed out that water 
pointed out that water and methanol were mainly produced by this method. 
In the close vicinity of a forming protostar, much higher temperature and higher 
fluxes of various energetic particles induce evaporation of the grain mantle.

Fascinating progress about the study of the interstellar molecules in ice phase started 
after the launch of Infrared Space Observatory (ISO) with the Short Wavelength Spectrometer 
(SWS) instrument on board. It was found that water is dominating the mantle composition 
and could account for $60-70$\% of the ice in most of the lines of sight 
\citep{whit03,gibb04}. Since icy mantles are mainly dominated by water 
molecules, abundances of any ice phase species mostly represented with respect to the ice phase water
abundance. Other detected ice phase species are $\mathrm{CO, \ CO_2, \ OCS, \ CH_3OH, \ CH_4}$ etc.. 
$\mathrm{OCN^-}$ \citep{whit01} and $\mathrm{NH_4^+}$ are supposed to be the 
carriers of some unidentified features as well. 

Composition of interstellar grain mantle  
varies throughout their evolutionary stages depending upon the amount of processing 
of the icy mantle materials. 
\citet{gibb04} classified various sources into five categories namely, Quiescent environment 
(example: Elias 16), Low-mass YSO (example: Elias 29), intermediate-mass YSO 
(example: AFGL 989, R CrA IRS 1), High-mass YSO with weak processing 
(example: W3 IRS 5, Mon R2 IRS 2, Mon R2 IRS 3) and High 
mass YSO with strong processing (example: W33A, AFGL 7009S). Around the quiescent dark molecular 
cloud regions, they found polar ($\mathrm{H_2O, \ CO}$ and possibly $\mathrm{NH_3}$ bearing) mantle 
coated by CO rich apolar mantle and trace amount of $\mathrm{CH_3OH}$ and $\mathrm{XCN}$. $\mathrm{CO}$ 
and $\mathrm{CO_2}$ are found to be covered by $\sim25$\% each with $\mathrm{NH_3}$ $<10$\%, 
Methanol $<3$\% and $\mathrm{XCN}$ $<1.5$\%. In case of Low-mass YSOs they found that $\mathrm{CO_2}$ 
is dominating ($\sim 20$\%), $\mathrm{CO}$ is $\sim 5$\%, $\mathrm{NH_3}$ is $<5$\%, $\mathrm{CH_3OH}$ 
is $<3$\% and $\mathrm{XCN}$ is $<1.5$\%. However, \citet{pont03} reported $\sim15$\% 
abundance of methanol in various low mass YSOs. This suggests that production of methanol is not 
only limited to high mass YSOs. Intermediate-mass YSOs are found to be dominated by 
large amounts of $\mathrm{CO_2}$ ($\sim30-35$\%) and $\mathrm{CO}$ ($18-50$\%) with trace amount 
of $NH_3$ ($<5$\%), methanol ($<5$\%) and $\mathrm{XCN}$ ($0.3$\%). High-mass YSOs with weak 
processing is found to be dominated by $\mathrm{CO_2}$ ($7-22$\%) and methanol ($<5-10$\%). 
Minor components are $\mathrm{CO}$ ($3-8$\%), $\mathrm{NH_3}$ ($<5$\%) and $\mathrm{XCN}$ ($0.3-2$\%). 
Around the High-mass YSOs with strong processing, ethanol, $\mathrm {CO_2, \ CO, \ NH_3}$ and 
$\mathrm{XCN}$ are found to be $15-30$\%, $13-23$\%, $8-17$\%, $15$\%, $2-6$\% respectively.

Despite low elemental abundances of atomic deuterium in interstellar space 
having average D/H ratio of $1.5 \times 10^{-5}$ \citep{lins95,lins03,stan98} in the ISM,
several species are found to be heavily fractionated \citep{wake14,das13b,das15b,maju14a,maju14b}.
In molecular clouds, the D reservoir is HD. The starting point 
of the deuterium fractionation is the following reaction between $\mathrm{{H_3}^+}$ and $\mathrm{HD}$ which 
produces $\mathrm{H_2D^+}$. This reaction can not follow the reverse step when kinetic temperature 
is less than $20$K. Thus abundances of $\mathrm{H_2D^+}$ steadily increase and $\mathrm{H_2D^+/{H_3}^+}$ 
ratio become significantly higher than the D elemental abundance with respect to H. CO and O are the 
major destruction partner of $\mathrm{H_2D^+}$. Around the CO and O depleted region, deuterium 
fractionation is further enhanced and produce multiply deuterated $\mathrm{H_3}^+$. 
Dissociative recombination of these multiply deuterated H3+ then produces D atoms and imply an atomic 
D/H ratio $>0.1$ \citep{case12}. This large atomic D/H ratio in gas phase 
effectively reflects major deuteration of the surface species. 
It is to be noted that extreme fractionation have only been observed in 
Low-mass prestellar cores and protostars. Probably due to the warmer environments, massive 
protostars do not show the same deuterium enrichment \citep{cecc07}.  

As per the elemental D/H ratio, singly deuterated species are expected to be $\sim 10^5$ 
times less abundant than their hydrogenated form. Similarly the doubly and triply 
deuterated forms are expected to be $\sim~10^{10}$ and $10^{15}$ times respectively lower. 
But in reality this is 
not be the case instead the deuterated ratios are 
found to be  extremely higher with enhancements of the D/H of up to $13$ orders of magnitude 
with respect to the elemental D/H abundance ratio \citep{case12}. 
Over the last few years multiply deuterated isotopologues of some common interstellar molecules, 
such as methanol, formaldehyde, thioformaldehyde, hydrogen sulfide and ammonia have been observed. 
Multiply deuterated formaldehydes have been observed around various sources. 
First doubly deuterated molecule detected in the ISM was D2CO. It was observed in Orion 
Compact Ridge (High-mas star forming regions) by \citet{turn90}. They found a fractionation value of 
$\sim 14\%$ for singly deuterated formaldehyde and $\sim 0.3\%$ for doubly deuterated formaldehyde. 
However, \citet{loin01} found a fractionation value $\sim 40 \pm 20\%$ 
for doubly deuterated formaldehyde towards the Low-mass 
protostar IRAS 16293E. Multiply deuterated ammonia molecules were 
observed during the last few years. Singly deuterated ammonia, $\mathrm{NH_2D}$ bears the fractionation 
ratio of $0.1$ in the class 0 source, NGC 1333 IRAS 4A \citep{vand00}. Doubly 
deuterated form of ammonia was observed by \citet{roue00} in a cold starless cloud, 
L134N having a fractionation ratio of $0.05$.  Moreover, triply deuterated ammonia, 
$\mathrm{ND_3}$ was detected by \citet{vand02} in NGC 1333 IRAS 4A with a fractionation 
ratio $0.001$. Deuterated methanols were detected along IRAS 16293 \citep{pari02,pari04}. 
They obtained the fractionation ratio for singly deuterated methanol 
$0.9 \pm 0.3$  and $0.04 \pm 0.02$ for $\mathrm{CH_2DOH/CH_3OH}$ and $\mathrm{CH_3OD/CH_3OH}$ 
respectively. In case of the 
doubly deuterated methanol $\mathrm{(CHD_2OH/CH_3OH})$, they derived a ratio of $0.2 \pm 0.1$.  
For the triply deuterated methanol ($\mathrm{CD_3OH/CH_3OH}$) they obtained a fractionation ratio averaged 
on a $10$ inch beam is $0.014$.

Water is found to be comparatively less fractionated in comparison to methanol around 
various sources. It was observed that the deuterium fractionation of water can be 
about $\mathrm{HDO/H_2O} \sim 0.01$\% towards massive hot cores \citep{gens96} and 
$\sim 0.02-0.05$\% 
in comets and asteroids \citep{altw03}.  According to the Vienna Standard 
Mean Ocean Water value, the fractionation ratio of $\mathrm{HDO/H_2O}$ is $\sim 1.588 \pm 0.001 \times 
10^{-4}$ in our oceans. However, some observations predicted a high fractionation ratio of a few \%
 towards the hot corino NGC1333-IRAS2A \citep{liu11}, and IRAS16283-2422 \citep{pari05}. 
\citet{butn07} discovered doubly deuterated water toward the protostellar binary 
system IRAS 16293-2422. They derived $\mathrm{D_2O/H_2O}$ ratio $5 \times 10^{-5}$ for 
the hot corino gas.  
Surface reactions play major role for the formation of water molecules in the ISM \citep{das11,caza10}. 
It is believed that it is most likely that the formation of water 
molecules on the surfaces of interstellar grains occurred before the formation of the solar nebula 
in the proto-solar molecular cloud \citep{jorg10,taqu13}. Isotopic composition of 
water along with the ortho-para ratio of the water formation epoch could be preserved in some 
objects like comets. Since ground base observation of water molecules are hindered by the huge 
abundance of water in the Earth’s atmosphere, various space based observations such as ISO, 
Spitzer, ODIN, SWAS, Herschel etc. were carried out. Deuterium fractionation of water could be 
used as an important diagnostics tool to find out a connection between the sources of Earth’s 
water fractionation (which is believed to have been brought in by the comets) and interstellar water.

\begin{table}
\centering
{\scriptsize
\caption{Initial gas phase abundances}
\begin{tabular}{|c|c|}
\hline
&{\bf Number density of cloud}\\
{\bf Species}&{\bf ($10^4$ cm$^{-3}$)}\\
\hline\hline
H&$1.1$\\
D&$1.1r_d$\\
O&$1.05$\\
CO&$0.15$\\
\hline
\end{tabular}}
\end{table}
In this paper, we mainly focus on the deuterium enrichment of interstellar grain mantle under various 
conditions. In Section 2, modeling details are presented. In Section 3, we present our 
results and finally in Section 4, we draw our conclusions.

\section{Modeling details}

\begin{table}
\centering
{\scriptsize
\caption{Surface Reactions considered}
\begin{tabular}{|l|l|l|}
\hline
& Reactions & E$_a$(K)            \\
\hline
1  & $\mathrm{H+H \rightarrow H_2}$          &      \\
2  & $\mathrm{H+O \rightarrow OH}$             &      \\
3  & $\mathrm{H+OH \rightarrow H_2O}$        &      \\
4  &$\mathrm{ H+CO \rightarrow HCO}$          & 390 \\
5  & $\mathrm{H+HCO \rightarrow H_2CO}$      &      \\
6  & $\mathrm{H+H_2CO \rightarrow H_3CO}$  & 415  \\
7  & $\mathrm{H+H_3CO \rightarrow CH_3OH}$ &      \\
8  & $\mathrm{O+O \rightarrow O_2}      $    &      \\
9 & $\mathrm{O+CO \rightarrow CO_2}     $   & 1000  \\
10 & $\mathrm{O+HCO \rightarrow CO_2+H} $    &      \\
11 & $\mathrm{O+O_2 \rightarrow O_3    } $    &       \\
12 & $\mathrm{H+O_2 \rightarrow HO_2    }$     & \\
13 & $\mathrm{H+HO_2 \rightarrow H_2O_2 }$        &  \\
14 & $\mathrm{H+H_2O_2 \rightarrow H_2O+OH}  $     &  \\
15 & $\mathrm{H+O_3 \rightarrow  O_2+OH    } $    &450  \\
16 & $\mathrm{H_2+OH \rightarrow H_2O+H}$&2600           \\
17 & $\mathrm{H+D \rightarrow HD}$          &      \\
18 & $\mathrm{D+D \rightarrow D_2}$          &      \\
19 & $\mathrm{D+O \rightarrow OD}$             &      \\
20 & $\mathrm{D+OH \rightarrow HDO}$        &      \\
21 & $\mathrm{H+OD \rightarrow HDO}$        &      \\
22 & $\mathrm{D+OD \rightarrow D_2O}$        &      \\
23 &$\mathrm{ D+CO \rightarrow DCO}$          & 320  \\
24 & $\mathrm{H+DCO \rightarrow HDCO}$      &      \\
25 & $\mathrm{D+HCO \rightarrow HDCO}$      &      \\
26 & $\mathrm{D+H_2CO \rightarrow H_2DCO}$  & 214  \\
27 & $\mathrm{D+HDCO \rightarrow HD_2CO}$  & 173  \\
28 & $\mathrm{D+D_2CO \rightarrow D_3CO}$  &  128  \\
29 & $\mathrm{H+HDCO \rightarrow H_2DCO}$  & 380  \\
30 & $\mathrm{H+D_2CO \rightarrow HD_2CO}$  & 340  \\
31 & $\mathrm{D+H_3CO \rightarrow CH_3OD}$ &      \\
32 & $\mathrm{D+H_2DCO \rightarrow CH_2DOD}$ &      \\
33 & $\mathrm{D+HD_2CO \rightarrow CHD_2OD}$ &      \\
34 & $\mathrm{D+D_3CO \rightarrow CD_3OD}$ &      \\
35 & $\mathrm{H+H_2DCO \rightarrow CH_3OD}$ &      \\
36 & $\mathrm{H+HD_2CO \rightarrow CH_2DOD}$ &      \\
37 & $\mathrm{H+D_3CO \rightarrow CHD_2OD}$ &      \\
38 & $\mathrm{O+DCO \rightarrow CO_2+D} $    &      \\
39 & $\mathrm{D+O_2 \rightarrow DO_2    }$     &   \\
40 & $\mathrm{D+HO_2 \rightarrow HDO_2 }$        &  \\
41 & $\mathrm{D+DO_2 \rightarrow D_2O_2 }$        &  \\
42 & $\mathrm{H+DO_2 \rightarrow HDO_2 }$        &  \\
43 & $\mathrm{D+O_3 \rightarrow  O_2+OD    } $    &450  \\
44 & $\mathrm{H_2+OD \rightarrow HDO+H}$&2600           \\
45 & $\mathrm{HD+OD \rightarrow HDO+D}$&2600           \\
46 & $\mathrm{HD+OD \rightarrow D_2O+H}$&2600           \\
47 & $\mathrm{HD+OH \rightarrow H_2O+D}$&2600           \\
48 & $\mathrm{D_2+OD \rightarrow D_2O+D}$&2600           \\
\hline
P1 & $\mathrm{O_2 \rightarrow O+O }$        &  \\
P2 & $\mathrm{OH  \rightarrow O+H }$        &  \\
P3 & $\mathrm{H_2O \rightarrow H+OH }$        &  \\
P4 & $\mathrm{CO \rightarrow C+O }$        &  \\
P5 & $\mathrm{HCO \rightarrow H+CO }$        &  \\
P6 & $\mathrm{H_2CO \rightarrow H_2+CO }$        &  \\
P7 & $\mathrm{CH_3OH \rightarrow CH_3+OH }$       &   \\
P8 & $\mathrm{CO_2 \rightarrow O+CO }$        &  \\
P9 & $\mathrm{H_2O_2 \rightarrow OH+OH }$       &   \\
P10 & $\mathrm{O_3 \rightarrow O_2+O }$        &  \\
P12 & $\mathrm{HO_2 \rightarrow OH+O }$        &  \\
P13 & $\mathrm{OD  \rightarrow O+D }$        &  \\
P14 & $\mathrm{D_2O \rightarrow D+OD }$        &  \\
P15 & $\mathrm{HDO \rightarrow H+OD }$         & \\
P16 & $\mathrm{DCO \rightarrow D+CO }$         & \\
P17 & $\mathrm{HDCO \rightarrow HD+CO }$        &  \\
P18 & $\mathrm{D_2CO \rightarrow D_2+CO }$        &  \\
P19 & $\mathrm{CH_3OD \rightarrow CH_3+OD }$        &  \\
P20 & $\mathrm{CH_2DOH \rightarrow CH_2D+OH }$        &  \\
P21  & $\mathrm{CHD_2OH \rightarrow CHD_2+OH }$        &  \\
P22 & $\mathrm{CH_2DOD \rightarrow CH_2D+OD }$        &  \\
P23 & $\mathrm{CD_3OH \rightarrow CD_3+OH }$        &  \\
P24 & $\mathrm{CHD_2OD \rightarrow CHD_2+OD }$        &  \\
P25 & $\mathrm{CD_3OD \rightarrow CD_3+OD }$        &  \\
P26 & $\mathrm{HDO_2 \rightarrow OD+OH }$        &  \\
P27 & $\mathrm{D_2O_2 \rightarrow OD+OD }$        &  \\
P28 & $\mathrm{DO_2 \rightarrow OD+O }$        &  \\
\hline
\end{tabular}}
\end{table}

\begin{figure}
\centering{
\includegraphics[height=8cm,width=8cm]{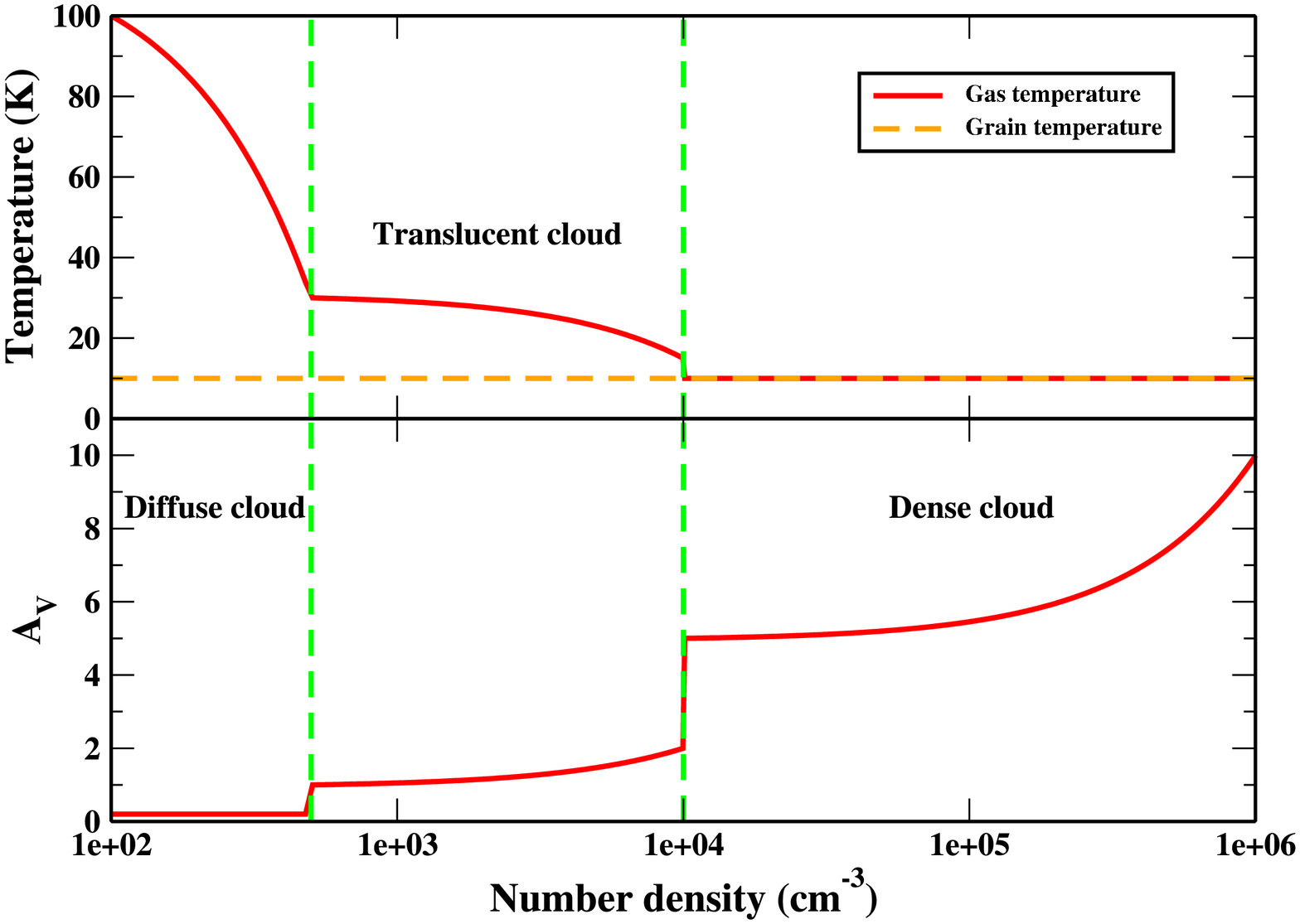}
\caption{Various types of clouds based on $n_H, T_{gas}$ and $A_V$.}
\label{fig-1}}
\end {figure}

\subsection{Chemical modeling}
It is well known that Monte Carlo method is computation intensive since Monte Carlo codes needs to handle
very different time-scales for different species. However, it gives the desired accuracy and the 
results are realistic. {Here, we use continuous time random walk Monte Carlo method to 
study the chemical evolution of the grain mantle. This code was originally developed by 
\citep{das10} and modified by \citet{das11}. In order to study the deuterium enrichment of the
grain mantle, we include some more reactions to this code.
To save computational time, we consider $50\times 50$ sites of grains
(as in \citet{das10,das11}) and extrapolate the results for the classical
grains (having $\sim 10^6$ sites). Reason for choosing
$50 \times 50$ site grains is that it is well above the limit of the statistical fluctuation
\citep{chan05}.} Here, all the events are processed by generating random numbers.
Both types of reaction mechanisms, namely, Langmuir-Hinshelwood 
(reactions between surface species via hopping) and Eley-Rideal 
(reaction between incoming species with surface species) are considered.

We assume that only gas phase $\mathrm{H,\ D,\ O}$ and $\mathrm{CO}$ are 
accreting on the grain surface. Unless otherwise stated, sticking probability ($s_n$) of all 
accreting species is assumed to be $1$. Initial gas phase number densities of these species 
are taken from the parameter space proposed by \citet{das10}. Initial abundances of these species for
$n_H=10^4$ cm$^{-3}$ are presented in Table 1. For the 
other density regions, initial abundances of these species are scaled accordingly. 
Initial abundance of gas phase atomic deuterium is assumed to be $r_d$ times the abundance 
of gas phase atomic hydrogen, where, $r_d$ is the initial atomic D/H ratio. Unless otherwise mentioned,
for most of our calculations, we use $r_d=0.3$ by following \citet{case02,osam04}.  
Since here we consider gas-grain interaction over the time, gas phase abundances of O and CO are 
considered to be depleted.

The reaction network considered here is shown in Table 2. Activation barriers are used by following \citet{das11} 
and references therein. These barriers are also shown in Table 2. For the deuterated reactions, 
activation barriers could differ due to zero point vibrations \citep{das15b,case02}. 
We follow \citet{case02} and updated activation barriers 
presented in \citet{fuch09} to update our activation barriers for the deuterated reactions.
As in \citet{das11}, here also, we consider some photo-reactions with similar
rate constants. Photo-reactions for the deuterated analogues are considered to have similar rate constants.

Binding energy of surface species controls the chemical composition of the
interstellar grain mantle. Normally, surface binding energies are computed from
various theoretical calculations (such as: \citet{alle77}). 
From experimental findings of \citet{pirr97,pirr99}, it was 
interpreted by \citet{katz99} that atomic hydrogen moves slower than what is usually assumed
in various simulations. Here, these experimental findings are also incorporated.
Since $\mathrm{D}$ atoms are heavier than $\mathrm{H}$ atoms, \citet{case02} used  
$2$ meV higher diffusion energy for $\mathrm{D}$ atom than $\mathrm{H}$ atom. 
Following this assumption, \citet{lips04}  
considered a $2$ meV difference between the diffusion energies of $\mathrm{H}$ and $\mathrm{D}$ atom.
\citet{lips04} also considered a $10$ meV energy difference between the desorption energies of
$\mathrm{H}$ and $\mathrm{D}$ atoms.
In Table 3, we show 4 Sets of binding energies which are considered in
our simulation. Since for all other species, we consider binding energies 
to be the same as in \citet{das10,das11}, only 
binding energies of $\mathrm{H, \  D, \ H_2, \ HD}$ and $\mathrm{D_2}$ are shown in the Table 3.
Set 1 corresponds to the theoretical values obtained by \citet{alle77}.
Unless otherwise stated, we always use Set 1 energy values. In Set 2 and Set 3, 
experimental findings \citep{katz99} of energies for olivine and amorphous carbon grains are 
respectively shown. Except $\mathrm{H, \ D, \, H_2, \ HD, \ D_2}$, all 
other energy values were kept unchanged as in Set 1.
Due to the unavailability of energy barriers, for Set 1, Set 2 and Set 3, 
binding energies of HD and $\mathrm{D_2}$ are assumed to be similar to 
$\mathrm{H_2}$. In Set 4, considerations of \citet{lips04} are taken care of. Since, no
energy barriers for $\mathrm{H_2}$, HD and $\mathrm{D_2}$ were available, we consider
$E_D(\mathrm {H_2})/E_D(\mathrm{H})$ ratios of energy values of Sets 2 and 3 for the computation of
$E_D(\mathrm{H_2})$ from $E_D(\mathrm{H})$ of \citet{lips04} case. A similar ratio 
($E_D(\mathrm{H_2})/E_D(\mathrm{H})$) is also used to compute $E_D(\mathrm{HD},
\mathrm{D_2})$ from $E_D(\mathrm{D})$ of \citet{lips04} case.

\begin{table*}
\centering{ 
\scriptsize
\caption{Different sets of binding energies}
\begin{tabular}{|c|c|c|c|c|c|c|c|c|}
\hline
{\bf Species}&\multicolumn{2}{|c|}{\bf Set 1}&\multicolumn{2}{|c|}{\bf Set 2}&\multicolumn{2}{|c|}{\bf Set 3} &\multicolumn{2}{|c|}{\bf Set 4}\\
&\multicolumn{2}{|c|}{\bf \citep{alle77}}&\multicolumn{2}{|c|}{\bf (\citet{katz99} for olivine)}&\multicolumn{2}{|c|}{\bf (\citet{katz99} for amorphous)} &\multicolumn{2}{|c|}{\bf \citep{lips04}}\\
\cline{2-9}\\
&${\bf E_b (K)}$&${\bf E_d (K)}$&${\bf E_b (K)}$&${\bf E_d (K)}$&${\bf E_b (K)}$&${\bf E_d (K)}$&${\bf E_b (K)}$&${\bf E_d (K)}$\\
\hline
\hline
$\mathrm{\bf H}$&100&350&287&373&511&657&406&580\\
$\mathrm{\bf D}$&100&350&287&373&511&657&429&697\\
$\mathrm{\bf H_2}$&135&450&94&314&163&542&143&478\\
$\mathrm{\bf HD}$&135&450&94&314&163&542&173&575\\
$\mathrm{\bf D_2}$&135&450&94&314&163&542&173&575\\
\hline
\end{tabular}}
\end{table*}

\subsection{Modeling the Physical Aspects}

The chemical composition of interstellar grain mantle solely depends on the surrounding 
physical conditions and age of the cloud. Presently, we wish to study chemical composition of interstellar
grain mantles for various types of clouds. \citet{snow06} classified clouds depending
on densities ($n_H$), visual extinctions ($A_V$) and temperatures ($T$). Here also, we are 
considering similar classification for our simulations.  For the diffuse cloud region, 
we consider that $n_H$ may vary between $10^2 - 5.00 \times 10^2$ cm$^{-3}$, $T$ may vary in 
between $30-100$ K and $A_V=0.2$, for the translucent cloud region, we use $n_H=5.01 \times 10^2- 
10^4$ cm$^{-3}$, $T=15-30$ K, $A_V=1-2$ and for the dense cloud region, we use $n_H=1.0001
\times 10^4 - 10^6$ 
cm$^{-3}$, $T=10$ K and $A_V=5-10$. In order to consider more realistic condition, we consider 
different temperatures and extinction parameters for various regions of clouds. We use constant 
slopes for $T$ and $A_V$ in respective number density windows. 
Fig. 1 clearly shows our choice of physical parameters. 
In diffuse and translucent clouds regions, gas and grains are 
not well coupled. So, temperature between these two phases might vary. However, in dense cloud regions
gas and grains are strongly coupled and the temperature would be more or less the same for both the phases. 
Here, we assume that the grain temperature ($T_{gr}$) is always fixed at $10$ K 
for all the clouds and gas temperature ($T_{gas}$), we assume $T_{gas}=T$ .

\section{Results}

\subsection{Composition of grain mantles with deuterated species}

Despite low elemental abundance of atomic deuterium in interstellar space, 
several species are found to be heavily fractionated in that sense, hydrogen atom is
replaced by deuterium. In order to find out the effects of this trace amount of deuterium on the
chemical composition, we study both the cases: without (Fig. 2a) and with 
deuterium (Fig. 2b, Fig. 2c, Fig. 2d and Fig. 2e).
Here, we consider $n_H=2 \times 10^4$ cm$^{-3}$ and corresponding values of $T_{gas}$ and 
$A_V$ are chosen according to Fig. 1. 
For Fig. 2a, 2b, Fig. 2c and Fig. 2d, we use $r_d=0,0.01,0.1$ and $0.3$ respectively and $s_n=1$. 
In case of Fig. 2e, we use $r_d=0.3$ but consider $s_n=0.3$ to see the effect
of sticking parameter on the composition of grain mantle.
The difference between Fig. 2d and Fig. 2e is clear. 
Consideration of the lower sticking probability largely affects the mantle composition. 
Formation of numbers of layers is significantly lower in case of Fig. 2e.
Fig. 2b, Fig 2c and Fig. 2d show the progressive increment of deuterium fractionation in comparison to
the non-deuterated case (Fig. 2a). For a better assessment, surface coverage, deuterium fractionation 
and abundances of some important surface species are shown in Table 4. Table 4 gives a 
comprehensive knowledge about various aspects of deuterium enrichment shown in Fig. 2(a-e). 
For simplicity, we are discussing only the case of $r_d=0.3$ and $s_n=1$ here.

\begin{table*}
\addtolength{\tabcolsep}{-5.3pt}
{\scriptsize
\caption{Surface coverage, deuterium fractionation and abundances of various surface species
with the absence and presence of deuterium.}
\begin{tabular}{|c|c|c|c|c|c|c|}
\hline
{\bf Species}&\multicolumn{5}{|c|}{\bf surface coverage/fractionation/abundance}&{\bf Observed abundance/}\\
&{\bf }& \multicolumn{4}{|c|}{\bf }&{\bf fractionation }\\
&{\bf $r_d=0, s_n=1$}&{\bf $r_d=0.01, s_n=1$}&{\bf $r_d=0.1, s_n=1$}&{\bf $r_d=0.3, s_n=1$}
&{\bf $r_d=0.3, s_n=0.3$}&{\bf }\\
\hline
\hline
$\mathrm{\bf H_2O}$&85.51/-/$9.28 \times 10^{-5}$&84.18/-/       9.15$\times 10^{-5}$&74.07/-/       8.09$\times 10^{-5}$&53.58/-/$5.92 \times 10^{-5}$&52.13/-/       3.57$\times 10^{-5}$&$4.7-40 \times 10^{-6}$$^c$/-\\
$\mathrm{\bf HDO}$&-/-/-&1.30/   0.02/       1.41$\times 10^{-6}$&11.21/   0.15/       1.22$\times 10^{-5}$&29.0/0.54/$3.20 \times 10^{-5}$&30.69/   0.59/       2.10$\times 10^{-5}$&$1.4-2.4 \times 10^{-7}/$\\
&&&&&&$(9.2\pm 2.6) \times 10^{-4}$$^a,$\\
&&&&&&$4-51 \times 10^{-3}$$^c$\\
$\mathrm{\bf D_2O}$&-/-/-&0.003/  0.00004/       3.72$\times 10^{-9}$&0.39/   0.01/       4.21$\times 10^{-7}$&3.40/0.06/$3.76 \times 10^{-6}$&4.02/   0.08/       2.75$\times 10^{-6}$&$<1.3 \times 10^{-9c}$/$<3 \times 10^{-4c}$\\
$\mathrm{\bf CH_3OH}$&7.75/-/$8.42 \times 10^{-6}$ &7.62/-/       8.28$\times 10^{-6}$& 6.47/-/       7.07$\times 10^{-6}$&4.66/-/$5.08 \times 10^{-6}$&3.30/-/       2.26$\times 10^{-6}$&\\
$\mathrm{\bf CH_3OD}$&-/-/-&0.09/   0.01/       9.31$\times 10^{-8}$&0.62/   0.10/       6.74$\times 10^{-7}$&1.25/0.27/$1.39 \times 10^{-6}$&0.91/   0.28/       6.24$\times 10^{-7}$&$0.02 (\pm 0.01)^b$\\
$\mathrm{\bf CH_2DOH}$&-/-/-&0.08/   0.01/       9.15$\times 10^{-8}$&0.63/   0.10/       6.83$\times 10^{-7}$&1.27/0.28/$1.41 \times 10^{-6}$&0.95/   0.29/       6.53$\times 10^{-7}$&$0.3 (\pm 0.2)^b$\\
$\mathrm{\bf CHD_2OH}$&-/-/-&0.002/   0.00026/       2.13$\times 10^{-9}$&0.09/   0.01/       1.03$\times 10^{-7}$& 0.51/0.110/$5.60 \times 10^{-7}$&0.37/   0.11/       2.56$\times 10^{-7}$&$0.06 (\pm 0.05)^b$\\
$\mathrm{\bf CH_2DOD}$&-/-/-&0.00098/   0.00013/       1.06$\times 10^{-9}$&0.10/   0.02/       1.07$\times 10^{-7}$&0.48/0.105/$5.34 \times 10^{-7}$&0.41/   0.13/       2.84$\times 10^{-7}$&-\\
$\mathrm{\bf CD_3OH}$&-/-/-&-/   -/       -&0.01/  0.002/       1.60$\times 10^{-8}$&0.17/0.04/$1.90 \times 10^{-7}$& 0.12/   0.04/       8.51$\times 10^{-8}$&$0.014(\pm 0.014)/0.008(\pm 0.006)^b$\\
$\mathrm{\bf CHD_2OD}$&-/-/-&-/   -/       -&0.01/   0.002/       1.28$\times 10^{-8}$&0.15/0.03/$1.68 \times 10^{-7}$& 0.12/   0.04/       8.09$\times 10^{-8}$&-\\
$\mathrm{\bf CD_3OD}$&-/-/-&-/   -/       -&-/   -/       -&0.04/0.008/$4.26 \times 10^{-8}$&0.02/   0.01/       1.33$\times 10^{-8}$&-\\
$\mathrm{\bf CO_2}$&4.34/-/$4.72 \times 10^{-6}$&4.27/-/       4.64$\times 10^{-6}$&3.99/-/       4.36$\times 10^{-6}$&3.31/-/$3.65 \times 10^{-6}$&4.28/-/       2.93$\times 10^{-6}$&\\
$\mathrm{\bf CO}$&0.07/-/$7.08\times 10^{-8}$&0.07/-/       7.55$\times 10^{-8}$&0.15/-/       1.65$\times 10^{-7}$&0.13/-/$1.45 \times 10^{-7}$& 0.04/-/       2.71$\times 10^{-8}$&\\
$\mathrm{\bf O_2}$&1.32/-/$1.43\times 10^{-6}$&1.30/-/       1.41$\times 10^{-6}$&1.09/-/       1.19$\times 10^{-6}$&0.73/-/$8.07 \times 10^{-7}$&1.04/-/       7.13$\times 10^{-7}$&\\
$\mathrm{\bf O_3}$&0.35/-/$3.83\times 10^{-8}$&0.04/-/       3.83$\times 10^{-8}$&0.02/-/       2.55$\times 10^{-8}$&0.02/-/$1.92 \times 10^{-8}$&0.03/-/       2.39$\times 10^{-8}$&\\
$\mathrm{\bf C}$&0.011/-/       1.17$\times 10^{-8}$&0.01/-/       8.51$\times 10^{-9}$&0.01/-/       1.17$\times 10^{-8}$&0.01/-/       1.12$\times 10^{-8}$&0.01/-/       7.98$\times 10^{-9}$&\\
$\mathrm{\bf OH}$&0.07/-/       7.55$\times 10^{-8}$&0.08/-/       8.19$\times 10^{-8}$&0.06/-/       6.06$\times 10^{-8}$&0.04/-/       4.26$\times 10^{-8}$&0.092/-/       6.33$\times 10^{-8}$&\\
$\mathrm{\bf OD}$&-/-/-&0.02/  0.30/       2.45$\times 10^{-8}$&0.17/  3.00/       1.82$\times 10^{-7}$&0.34/  8.80/       3.75$\times 10^{-7}$&0.47/  5.05/       3.20$\times 10^{-7}$&\\
$\mathrm{\bf CH_3}$&0.86/-/       9.29$\times 10^{-7}$&0.90/-/       9.80$\times 10^{-7}$&0.77/-/       8.38$\times 10^{-7}$&0.66/-/       7.25$\times 10^{-7}$&0.64/-/       4.37$\times 10^{-7}$&\\
$\mathrm{\bf CH_2D}$&-/-/-&0.01/  0.01/       1.06$\times 10^{-8}$&0.07/  0.097/       8.14$\times 10^{-8}$&0.18/  0.27/       1.95$\times 10^{-7}$&0.21/  0.33/       1.44$\times 10^{-7}$&\\
\hline
{\bf No. of}&85&85&86&87&55&\\
{\bf mono-layers}&&&&&&\\
\hline
\multicolumn{6}{|c|}{$^a$ \citet{pers12} (along IRAS 16293-2422)} \\
\multicolumn{6}{|c|}{$^b$ \citet{pari04} (along IRAS 16293-2422)}\\
\multicolumn{6}{|c|}{$^c$ \citet{cout13} (along hot corino of IRAS 16293)}\\
\hline
\end{tabular}}
\vskip 1cm
\end{table*}

From Fig. 2a ($r_d=0$), we find that water and methanol are maintaining surface
coverages of $85.5\%$ and $7.8\%$ respectively. In case of Fig. 2d, where $r_d=0.3$  is considered, 
surface coverage of water in all forms and methanol in all forms come out to be $86$\% (normal 
water is $53.6\%$ and deuterated water is $32.4\%$)  and $8.5\%$ (normal methanol is $4.6\%$ and 
deuterated methanol is $2.9\%$) respectively. Among the deuterated species, $\mathrm{HDO}$ is 
mostly dominating (surface coverage $29\%$) while $\mathrm{D_2O}$ is also found to be abundant 
(surface coverage $3.4\%$). 
Deuterium fractionation of singly deuterated methanols, i.e., for
$\mathrm{CH_3OD}$ and $\mathrm{CH_2DOH}$ are found to be
$0.27$ and $0.28$ respectively and seems to occupy significant percentage of the grain mantle ($1.25\%$ and $1.27\%$ respectively). For a similar choice of $r_d$, fractionation values for the
doubly deuterated methanols are found to be $0.110$ and $0.105$ respectively for $\mathrm{CHD_2OH}$ and 
$\mathrm{CH_2DOD}$. For a similar situation, triply deuterated methanols possess 
fractionation ratios of $0.04$ and $0.03$ respectively for $\mathrm{CD_3OH}$ and $\mathrm{CHD_2OD}$. 
For $\mathrm{CD_3OD}$, this ratio is found to be $0.008$. 

Triply deuterated methanol ($\mathrm{CD_3OH}$) has recently been observed to have a fractionation ratio of 
$0.014 (\pm 0.014)/0.008 ( \pm 0.006) $) in IRAS 16293-2422 by \citet{pari04}. IRAS 16293-2422 is 
a Class 0 protostellar binary. It is expected that due to the heating of newly formed stars, 
ice mantles are evaporated and 
grain phase species are released into the gas phase \citep{char92,char97,case93,pari04}. 
One would expect that deuterium fractionation of 
the ice phase would be preserved in the gas phase as well but this is not true for all
the species. In Table 4, we compare our calculated water and methanol abundance/fractionation 
with existing observational results. Doubly deuterated methanol 
($\mathrm{CHD_2OH}$) is also observed by \citet{pari04} having a fractionation ratio of 
$0.06 (\pm 0.05)$. Our calculated results for $r_d=0.3$ is in good agreement for the fractionation ratios 
for doubly and triply deuterated methanol. In case of singly deuterated methanols, we are having 
roughly similar fractionation ratios. Our obtained result for $\mathrm{CH_2DOH}$ is very close to 
the observed value. But observational evidences suggest that
$\mathrm{CH_3OD}$ is relatively less abundant in comparison to  
$\mathrm{CH_2DOH}$ and fractionation ratio of $\mathrm{CH_3OD/CH_3OH}$
and $\mathrm{CH_2DOH/CH_3OH}$ would be $0.02 (\pm 0.01)$ and $0.3 (\pm 0.2)$ 
respectively in IRAS 16293-2422. 
\citet{char97} and \citet{osam04} discussed the possible 
interconversion between pairs of deuterated forms of methanol, its ion, and its 
proportionated ion in star-forming regions. They also suggested that $\mathrm{CH_3OD}$ is comparatively 
less abundant than $\mathrm{CH_2DOH}$. \citet{osam04} carried out quantum chemical calculations 
to show various energy required for the possible internal rearrangements. 
They assumed that methanol along with its deuterated forms mainly form on 
interstellar dust and are evaporated into gas phase by several other means and 
the interconversion mainly take place after the evaporation from the dusts. 
They ran some protostellar models, starting immediately after the loss of grain mantles. 
Here, we consider only depletions of some gas phase species. Evaporation of 
surface species is considered by cosmic ray induced process and thermal process only. Since
we consider grain temperature to be fixed at $10$ K, in our case, no such bulk evaporation takes 
place. Consideration of detailed gas phase chemistry along with the Monte Carlo 
simulation is also out of scope for this paper. Thus the fractionation ratios obtained from our 
simulation only reflects the values in the cold phase only. 
This ratio might be changed after the evaporation of grain mantles because of the possible interaction 
with some dominant gas phase species. \citet{case02} also studied the formation of deuterated isotopomers of 
methanol on interstellar grain surfaces with a semi-empirical modified rate approach and a 
Monte Carlo method in between the temperature range of $10-20$ K. For intermediate density
cloud ($n_H=10^4$ cm$^{-3}$), their Monte Carlo simulation showed that the fractionation ratios 
after $10^4$ year with an initial atomic $\mathrm{D/H}$ ratio of $0.3$ are 
$0.19,0.19, 3.3 \times 10^{-4}, 0.034$ and $6.6 \times 10^{-5}$ respectively for 
$\mathrm{CH_3OD, \ CH_2DOH, \ CHD_2OH, \ CHD_2OD}$ and $\mathrm{CHD_2OD}$. Since they also did not
follow the time evolution of the gas phase species after the bulk evaporation, 
they also obtained similar fractionation 
ratios for the singly deuterated methanols.

Though a high level of deuterium fractionation of methanol 
is evident from the observations, surprisingly, a low fractionation ratio is observed for $\mathrm{HDO}$. 
This is in contradiction with our results. Origin of this discrepancy is yet to be understood and could indicate new processes not incorporated in our study.
\citet{pers12} measured deuterium fractionation 
in the warm gas of the deeply-embedded protostellar binary IRAS 16293-2422. They found a 
fractionation ratio of $9.2 \pm 2.6 \times 10^{-4}$ for solid $\mathrm{HDO}$. \citet{pers12} 
predicted an upper limit of $10^{-2}$ for $\mathrm{HDO/H_2O}$ in NGC 7358 IRS9. 
\citet{pari03} discusses various scenarios which could explain such low water deuteration 
compared to the observed high deuteration of methanol in the gas-phase.
\citet{wake14} model the chemistry occurring on IRAS 16293-2422 with a large gas-grain chemical
code. They used various physical conditions to reproduce the observed 
$\mathrm{HDO/H_2O}$ and $\mathrm{D_2O/H_2O}$. They found that  
by considering a cosmic ray ionization rate of $10^{-16}$ S$^{-1}$ and C/O elemental
ratio of $0.5$, observed ratios would be reproduced around the central part of the 
protostellar envelope but their model was not able to reproduce the observed ratios for 
the outer part of the protostellar envelope. This important unresolved puzzle clearly requires further investigation.

\subsection{Time evolution of mantle composition}

Results presented in Fig. 2(a-e) and Table 4 represent the status after $2 \times 10^6$ years. 
We call  this time as the 'standard time'. Fig. 2(a-e) shows that several mono-layers would be 
produced during the simulation period. Studies by Flower et al. (2006) suggested that evolution 
of layers can take place even in longer time scales. In order to see the effects of time on 
the formation of number of mono-layers, we carry out our simulation for much longer 
time scales ($~10^7$ years). Fig. 3 shows the formation of number of layers in different time scales. 
'Early time' corresponds to $10^4$ years, 'intermediate time' corresponds to $10^5$ years and 
'late time' corresponds to $10^7$ years. By the number of layers, we mean the maximum number of 
layers counting from the bare grain surface on at least one site. Thus, $n^{th}$ layer starts with the 
first instance of deposition of any species on any site which already has (n-1) layer. 
From Fig. 3, it is clear that the formation of number of layers is highly time dependent phenomenon. 
For the low density region, formation of number of layers is linearly increasing with time. 
Situation is somewhat different in the high density region. For the high density region, i.e., 
when $n_H > 2 \times 10^4$ cm$^{-3}$ region, hardly a few new layers are produced beyond standard time. 
Thus it is not useful to continue simulation for longer period. Thus to save computational time, 
we computed only up to the standard time and showed our results.

This aspect would be better understood from Fig. 4(a-c), where chemical evolution of the 
most dominating gas-grain species of our chemical network is shown for $n_H=10^3$,  $2 \times 10^4$ \& $10^6$ 
cm$^{-3}$ cloud. Corresponding values of the temperatures and extinction parameters are chosen from Fig. 1. 
Along the 'y' axis of Fig. 4(a-c), we show the abundances of gas-grain species with respect to 
total hydrogen nuclei in all form and `x' axis shows time in years. For this case, we continue 
our simulation up to $10^7$ years. For the low density region ($n_H=10^3$ cm$^{-3}$), 
steady state 
between the surface species is not achieved. For the higher density regions ($n_H=2 \times 10^4$ and $10^6$ 
cm$^{-3}$)
, a steady state among the surface species would arise within the standard time. This is possible 
due to depletion of gas phase species.

For Fig. 4a (low density), we use $n_H=10^3$ cm$^{-3}$ and strong radiation field ($A_V \sim 1.05$). 
In this low density region, formation of complex species is mainly countered by the photo-dissociation 
process. Due to this reason, photo dissociation products like $\mathrm{C,\ CH_3,\ CH_2D,\ OH,\ OD}$ are 
covering significant parts of the grain mantles. For the high density regions, gas phase $\mathrm{O}$ and 
$\mathrm{CO}$ are heavily depleted and steady state is achieved in much shorter time scales. 
Both the deuterated forms of $\mathrm{H_2O}$ ($\mathrm{HDO}$ \& $\mathrm{D_2O}$) are produced 
efficiently when $n_H=2 x 10^4$ cm$^{-3}$. Singly deuterated forms of methanol ($\mathrm{CH_3OD}$ and 
$\mathrm{CH_2DOH}$) are also found to be covering a significant percentage of grain mantles. $\mathrm{O_3}$
 is found to be the most dominating species in the much higher density region ($10^6$). 
Fig. 4(a-c) clearly shows how time affects the abundances of gas-grain  species for the low density region. 
However, for the sake of benchmarking, we need to choose a time scale to compare various 
aspects of grain mantles.

\subsection{Analysis of the grain mantle composition in different types of clouds}

\begin{table*}
{\scriptsize
\caption{Surface coverage, fractionation ratio and abundance of some important surface species for various types of clouds.}
\begin{tabular}{|c|c|c|c|c|c|}
\hline
{\bf Species}&\multicolumn{5}{|c|}{\bf Surface coverage/deuterium fractionation/ abundance}\\
& \multicolumn{2}{|c|}{\bf Dense cloud}& \multicolumn{2}{|c|}{\bf Translucent cloud}&{\bf Diffuse cloud} \\
& {\bf $n_H=10^6$ cm$^{-3}$}& {\bf $n_H=2 \times 10^4$ cm$^{-3}$}& {\bf $n_H=5  \times 10^3$ cm$^{-3}$} & {\bf $n_H= 10^3$ cm$^{-3}$}&{\bf $10^2$ cm$^{-3}$} \\
\hline
\hline
$\mathrm{\bf H_2O}$&11.23/-/       6.55$\times 10^{-06}$&53.58/-/       5.92$\times 10^{-05}$& 14.06/-/       7.79$\times 10^{-06}$&21.27/-/       1.57$\times 10^{-06}$&20.16/-/       2.48$\times 10^{-07}$\\
$\mathrm{\bf HDO}$& 4.77/   0.43/       2.78$\times 10^{-06}$&29.00/   0.54/       3.20$\times 10^{-05}$&28.91/   2.06/       1.60$\times 10^{-05}$&20.05/   0.94/       1.48$\times 10^{-06}$&22.50/   1.12/       2.77$\times 10^{-07}$\\
$\mathrm{\bf D_2O}$&0.62/   0.06/       3.61$\times 10^{-07}$&3.40/   0.06/       3.76$\times 10^{-06}$&7.02/   0.50/       3.89$\times 10^{-06}$&3.90/   0.18/       2.87$\times 10^{-07}$&3.38/   0.17/       4.15$\times 10^{-08}$\\
$\mathrm{\bf CH_3OH}$&0.88/-/       5.15$\times 10^{-07}$&4.60/-/       5.08$\times 10^{-06}$&0.11/-/       5.85$\times 10^{-08}$& 0.04/-/       3.19$\times 10^{-09}$&-/-/       -\\
$\mathrm{\bf CH_3OD}$&0.31/   0.36/       1.83$\times 10^{-07}$&1.25/   0.27/       1.39$\times 10^{-06}$&0.02/   0.22/       1.28$\times 10^{-08}$&-/   -/       -&-/-/       -\\
$\mathrm{\bf CH_2DOH}$&0.36/   0.41/       2.12$\times 10^{-07}$&1.27/   0.28/       1.41$\times 10^{-06}$&0.03/   0.33/       1.92$\times 10^{-08}$&-/   -/       -&-/-/       -\\
$\mathrm{\bf CHD_2OH}$&0.20/   0.23/       1.18$\times 10^{-07}$&0.51/   0.11/       5.60$\times 10^{-07}$&0.01/   0.09/       5.32$\times 10^{-09}$&0.01/   0.17/       5.32$\times 10^{-10}$&-/-/       -\\
$\mathrm{\bf CH_2DOD}$&  0.24/   0.27/       1.39$\times 10^{-07}$&0.48/   0.11/       5.34$\times 10^{-07}$&0.02/   0.16/       9.58$\times 10^{-09}$&-/   -/       -&-/-/       -\\
$\mathrm{\bf CD_3OH}$&0.09/   0.10/       5.27$\times 10^{-08}$&0.17/   0.04/       1.90$\times 10^{-07}$&-/   0.04/       2.13$\times 10^{-09}$& -/   -/       -&-/-/       -\\
$\mathrm{\bf CHD_2OD}$&0.08/   0.09/       4.42$\times 10^{-08}$&0.15/   0.03/       1.68$\times 10^{-07}$&-/   0.02/       1.06$\times 10^{-09}$& -/   -/       -&-/-/       -\\
$\mathrm{\bf CD_3OD}$& 0.03/   0.03/       1.49$\times 10^{-08}$&0.04/   0.01/       4.26$\times 10^{-08}$&-/   0.01/       5.32$\times 10^{-10}$& -/   -/       -&-/-/       -\\
$\mathrm{\bf CO_2}$&5.66/-/       3.30$\times 10^{-06}$&3.31/-/       3.65$\times 10^{-06}$&0.12/-/       6.38$\times 10^{-08}$&0.01/-/       1.06$\times 10^{-09}$& -/-/       -\\
$\mathrm{\bf CO}$&15.20/-/       8.86$\times 10^{-06}$&0.13/-/       1.45$\times 10^{-07}$&0.03/-/       1.76$\times 10^{-08}$&0.04/-/       2.66$\times 10^{-09}$&-/-/       -\\
$\mathrm{\bf O_2}$& 5.05/-/       2.94$\times 10^{-06}$&0.73/-/       8.07$\times 10^{-07}$&4.72/-/       2.61$\times 10^{-06}$&0.81/-/       5.96$\times 10^{-08}$&0.22/-/       2.66$\times 10^{-09}$\\
$\mathrm{\bf O_3}$&47.90/-/       2.79$\times 10^{-05}$&0.02/-/       1.92$\times 10^{-08}$&13.09/-/       7.25$\times 10^{-06}$&3.22/-/       2.37$\times 10^{-07}$&0.35/-/       4.26$\times 10^{-09}$\\
$\mathrm{\bf C}$&0.03/-/       2.02$\times 10^{-8}$&0.01/-/       1.12$\times 10^{-8}$&10.61/-/       6.46$\times 10^{-6}$&32.89/-/       2.42$\times 10^{-6}$&31.93/-/       3.93$\times 10^{-7}$\\
$\mathrm{\bf OH}$&2.28/-/       1.33$\times 10^{-6}$&0.04/-/       4.26$\times 10^{-8}$&2.77/-/       1.69$\times 10^{-6}$& 2.87/-/       2.11$\times 10^{-7}$& 1.87/-/       2.29$\times 10^{-8}$\\
$\mathrm{\bf OD}$&0.67/  0.29360/       3.90$\times 10^{-7}$&0.34/  8.80/       3.75$\times 10^{-7}$&6.70/  2.42/       4.08$\times 10^{-6}$&4.65/  1.62/       3.42$\times 10^{-7}$& 2.08/  1.12/       2.55$\times 10^{-8}$\\
$\mathrm{\bf CH_3}$&0.64/-/       3.71$\times 10^{-7}$&0.66/-/       7.25$\times 10^{-7}$& 8.58/-/       5.23$\times 10^{-6}$& 6.48/-/       4.77$\times 10^{-7}$& 12.03/-/       1.48$\times 10^{-7}$\\
$\mathrm{\bf CH_2D}$&0.13/  0.21/       7.71$\times 10^{-8}$&0.18/  0.27/       1.95$\times 10^{-7}$& 1.38/  0.16/       8.40$\times 10^{-7}$&0.82/  0.13/       6.06$\times 10^{-8}$& 1.56/  0.13/       1.92$\times 10^{-8}$\\
\hline
{\bf No. of}&49&87&50&13&6\\
{\bf mono-layers}&&&&&\\
\hline
\end{tabular}}
\vskip 2cm
\end{table*}

In Fig. 5(a-e), we show the composition of grain mantles for various cloud regions.
To represent diffuse cloud region, we use $n_H=10^2$ cm$^{-3}$ (Fig. 5a). For a translucent
cloud, we use $n_H=10^3$ cm$^{-3}$ (Fig. 5b) and $n_H=5 \times 10^3$ cm$^{-3}$ (Fig. 5c). 
To represent a dense cloud region, we use $n_H=2 \times 10^4$ cm$^{-3}$ (Fig. 5d) and
$n_H=10^6$ cm$^{-3}$ (Fig. 5e). Respective values of $T_{gas}$, $T_{gr}$, $A_V$ are chosen according 
to the physical profiles described in Fig. 1. Mantle compositions are plotted after a
standard time. Surface coverage, fractionation and abundances of some
important surface species are given in Table 5. 

In diffuse cloud regions (Fig. 5a), the radiation field is very strong. So hardly a few layers would
be produced. In translucent cloud regions (Fig. 5b and Fig. 5c), the radiation field is somewhat attenuated but 
its effect is till there. Thus the surface coverage is mainly dominated by the photo-dissociation
products  such as $\mathrm{C,CH_3,CH_2D,OH,OD}$ etc. Formation of number of layers is progressively 
higher for when the density is increased from $10^3$ cm$^{-3}$ to $5 \times 10^3$ cm$^{-3}$.
For a higher density region (Fig. 5d and Fig. 5e), radiation field is almost attenuated. Thus,
formation of complex molecules are  favoured. One interesting fact could be noted from Table 5 
that as we are going to the higher density side of the dense cloud region 
(from $n_H=2 \times 10^4$ cm$^{-3}$ to $10^6$ cm$^{-3}$), surface coverage of $\mathrm{H_2O}$ 
and $\mathrm{CH_3OH}$ along with its deuterated forms drastically drops. 
In general, abundances of methanol and water increases with density and attain a maximum 
around $n_H=2 \times 10^4$ cm$^{-3}$. If we further increase the 
density beyond $n_H=2 \times 10^4$ cm$^{-3}$,
their abundance starts to drop. So there are actually two regions, one region is extended from 
$10^2$ cm$^{-3}$ to $2 \times 10^4$ cm$^{-3}$ and another is beyond $n_H=2 \times 10^4$ cm$^{-3}$. 
From Table 5 it is evident that water abundance which was $2.48 \times 10^{-7}$ in $n_H=10^2$ cm$^{-3}$ 
region, increases to $5.92 \times 10^{-5}$ in $n_H=2 \times 10^4$ cm$^{-3}$ region. Beyond 
$n_H=2 \times 10^4$ cm$^{-3}$, it starts to drop and at $n_H=10^6$ cm$^{-3}$ it comes out to be 
$6.55 \times 10^{-6}$. Similar is also true for methanol. Production of methanol for the 
diffuse cloud region ($n_H=10^2$ cm$^{-3}$) was not significant but its production 
steadily increase with the increase in density and its abundance comes out to be 
$5.08 \times 10^6$ in $n_H=2 \times 10^4$ cm$^{-3}$. Beyond $2 \times 10^4$ cm$^{-3}$, 
methanol abundance starts to decrease and finally comes out to be $5.15 \times 10^{-7}$ for $n_H=10^6$ 
cm$^{-3}$.  
On the contrary, around the second region ($n_H=2 \times 10^4$ cm$^{-3}$ to $10^6$ cm$^{-3}$),
surface coverage of $\mathrm{CO, \ CO2, \ O_2, \ O_3}$ significantly increases. 
Since for $2 \times 10^4$ cm$^{-3}$ accretion rate is comparatively lower, surface species gets 
adequate time to recombine and produce various hydrogenated or deuterated species. 
This region of cloud is found to be favourable for the production of maximum numbers of species and 
thus forming maximum number of layers. Fig. 3 also shows that beyond the standard time, the region with
$n_H=2 \times 10^4$ cm$^{-3}$ produces maximum number of layers.
In this region, most of the incoming H and D atoms are utilized to form new molecules because
they get adequate time for the reactions. Thus this region produces more numbers of 
molecules and more numbers of layers. For the higher density side, accretion rate is 
high and H and D would be blocked by any unwanted species next to them. Thus H and D would 
evaporate from that site without making any reaction. Thus less number of species 
would be produced and less number of layers would be produced when $n_H>2 \times 10^4$ cm$^{-3}$. 
For the low density region, since the accretion rate is comparatively lower, it is not expected to produce 
more layers within the specified time. As we are increasing the density (diffuse to dense), 
Fig. 3 shows an increasing profile for the standard time. However, for $n_H>2 \times 10^4$ cm$^{-3}$,
numbers of layers are decreasing. For example, in case of $n_H=10^6$ cm$^{-3}$ region, 
comparatively less numbers of layers are produced. 
Here, the surface is full of various species, so $\mathrm{H}$ or $\mathrm{D}$ atoms could
be easily blocked by other unfavourable species (reactions either  not favourable 
or possess high activation barrier) and
are thermally evaporated after their residence time. Thus formation of hydrogenated or deuterated
species are heavily hindered. As a result, $\mathrm{CO}$ mostly remain unutilized on the grain. 
Atomic oxygens, which are mostly converted to $\mathrm{H_2O}$ and its deuterated forms in case
of $n_H=2 \times 10^4$ cm$^{-3}$ region, remain unutilized here and channelized to form 
$\mathrm{O_2, \ O_3}$ and $\mathrm{CO_2}$. For better illustration  purpose, in Fig. 6, 
we show abundances of various surface species in relation to number density.
Fig. 6 clearly shows that water and methanol abundances steadily increase
with the increase in $n_H$. However, when $n_H > 2 \times 10^4$ cm$^{-3}$, they are significantly decreased. 
Throughout the regions, abundances of $CO$ is found to be linearly increasing with 
the number density of clouds. Abundance of $CO_2$ is found to increase with the increase 
in $n_H$ but beyond $n_H=2 \times 10^4$ cm$^{-3}$, it roughly maintains a steady state. 
Photo-dissociation products which are abundant around the diffuse and translucent cloud regions 
are sharply decreasing in regions with $n_H > 5 \times 10^3$ cm$^{-3}$. 


\subsection{Composition of grain mantles for various sets of binding energies}

\begin{table*}
\addtolength{\tabcolsep}{-2pt}
\centering{ 
\scriptsize
\caption{Surface coverage, deuterium fractionation ratio and abundance for various sets of binding energies 
for $n_H=2 \times 10^4$ cm$^{-3}$}
\begin{tabular}{|c|c|c|c|c|}
\hline
{\bf Species}&\multicolumn{4}{|c|}{\bf surface coverage/ deuterium fractionation ratio/abundance}\\
&{\bf Set 1}&{\bf Set 2}&{\bf Set 3}&{\bf Set 4}\\
\hline
\hline
$\mathrm{\bf H_2O}$&53.58/-/  5.92 $\times 10^{-5}$& 41.12/-/  4.15$\times 10^{-05}$ & 25.39/-/  1.64$\times 10^{-05}$& 25.17/-/  1.61$\times 10^{-05}$\\
$\mathrm{\bf HDO}$&29.00/  0.54/  3.20 $\times 10^{-5}$& 33.14/  0.81/  3.35$\times 10^{-05}$ & 12.49/  0.49/  8.04$\times 10^{-06}$& 12.37/  0.49/  7.93$\times 10^{-06}$\\
$\mathrm{\bf D_2O}$&3.40/  0.06/  3.76 $\times 10^{-6}$&  6.38/  0.16/  6.44$\times 10^{-06}$& 2.00/  0.08/  1.28$\times 10^{-06}$ & 2.03/  0.08/1.30$\times 10^{-06}$\\
$\mathrm{\bf CH_3OH}$& 4.60/-/  5.08$\times 10^{-6}$& 2.13/-/  2.15$\times 10^{-06}$ & 0.09/-/  6.06$\times 10^{-08}$& 0.08/-/  5.32$\times 10^{-08}$\\
$\mathrm{\bf CH_3OD}$& 1.25/  0.27/  1.39 $\times 10^{-6}$& 0.41/  0.19/  4.10$\times 10^{-07}$ & 0.012/  0.13/  7.98$\times 10^{-09}$& 0.009/  0.11/  5.85$\times 10^{-09}$\\
$\mathrm{\bf CH_2DOH}$&1.27/  0.28/  1.41 $\times 10^{-6}$& 0.40/  0.19/  4.06$\times 10^{-07}$ & 0.006/  0.06/  3.72$\times 10^{-09}$& 0.008/  0.1/  5.32$\times 10^{-09}$\\
$\mathrm{\bf CHD_2OH}$& 0.51/  0.59/5.6 $\times 10^{-7}$& 0.08/  0.04/  7.66$\times 10^{-08}$ & 0.0017/  0.018/  1.06$\times 10^{-09}$ & 0.002/  0.02/  1.064$\times 10^{-09}$\\
$\mathrm{\bf CH_2DOD}$& 0.48/  0.105/  5.34$\times 10^{-7}$& 0.07/  0.04/  7.55$\times 10^{-08}$ & 0.002/  0.02/  1.06$\times 10^{-09}$& 0.0008/  0.01/  5.32$\times 10^{-10}$\\
$\mathrm{\bf CD_3OH}$& 0.17/  0.04/  1.90$\times 10^{-7}$& 0.04/  0.02/  3.56$\times 10^{-05}$ & -/  - /  -                  & - /  - /  -                  \\
$\mathrm{\bf CHD_2OD}$& 0.15/  0.03/  1.68$\times 10^{-7}$& 0.03/  0.01/  2.87$\times 10^{-05}$ & - /  - /  -                  & - /  - /  -                  \\
$\mathrm{\bf CD_3OD}$& 0.04/  0.008/  4.26$\times 10^{-8}$& 0.01/  0.005/  1.06$\times 10^{-08}$ & -/-/ -                 & - /  -/  -                  \\
$\mathrm{\bf CO_2}$& 3.31/-/  3.65$\times 10^{-6}$& 3.06/-/  3.09$\times 10^{-06}$&0.66/-/  4.27 $\times 10^{-07}$ & 0.67/-/  4.30$\times 10^{-07}$\\
$\mathrm{\bf CO}$&  0.13/-/  1.45$\times 10^{-7}$& 4.91/-/  4.95$\times 10^{-06}$& 20.05/-/  1.29$\times 10^{-05}$ &19.93/-/  1.27$\times 10^{-05}$\\
$\mathrm{\bf O_2}$& 0.73/-/  8.07$\times 10^{-7}$& 0.46/-/  4.64$\times 10^{-07}$& 0.07/-/  4.68$\times 10^{-08}$&0.07/-/  4.52 $\times 10^{-08}$\\
$\mathrm{\bf O_3}$& 0.02/-/  1.92$\times 10^{-8}$& 4.89/-/  4.93$\times 10^{-06}$ & 38.22/-//2.46$\times 10^{-05}$& 38.44/-/  2.46$\times 10^{-05}$\\
$\mathrm{\bf C}$&0.01/-/       1.12$\times 10^{-8}$&0.03/-/       3.51$\times 10^{-8}$&70.007/-/       4.26$\times 10^{-9}$&0.12/-/       7.55$\times 10^{-8}$\\
$\mathrm{\bf OH}$&0.04/-/       4.26$\times 10^{-8}$&0.04/-/       3.62$\times 10^{-8}$&0.11/-/       7.24$\times 10^{-8}$&0.18/-/       1.17$\times 10^{-7}$\\
$\mathrm{\bf OD}$&0.34/  8.80/       3.75$\times 10^{-7}$&0.34/  9.44/       3.42$\times 10^{-7}$&0.31/  2.78/       2.01$\times 10^{-7}$&0.34/  1.89/       2.20$\times 10^{-7}$\\
$\mathrm{\bf CH_3}$&0.66/-/       7.25$\times 10^{-7}$&0.25/-/       2.57$\times 10^{-7}$&-/-/-&0.01/-/       9.04$\times 10^{-9}$\\
$\mathrm{\bf CH_2D}$&0.18/  0.27/       1.95$\times 10^{-7}$&0.04/  0.18/       4.52$\times 10^{-8}$&-/-/-&0.002/  0.12/       1.06$\times 10^{-9}$\\
\hline
{\bf No. of}&87&80&52&53\\
{\bf mono-layers}&&&&\\
\hline
\end{tabular}}
\end{table*}

Till now, we used Set 1 energy barriers. Presently, we consider all $4$ sets of energy barriers 
given in Table 3. Chemical composition of the grain mantles under these energy barriers 
is shown in Fig. 7(a-d). In this case, we use
$n_H=2 \times 10^4$ cm$^{-3}$ and choose $A_V,T_{gas}, T_{gr}$ from Fig. 1. For quantitative
comparison, we show the surface coverage, deuterium fractionation and abundances of important 
surface species in Table 6. It is clear from Fig. 7(a-d) and Table 6 that methanol is efficiently 
produced for Set 1 binding energies. Hydrogenation reactions are the fastest     
reactions on the grain surface. Since for Set 1, energy barrier against diffusion is 
the lowest for atomic hydrogen (see, Table 3 for the energy barriers), hydrogenation reactions 
are frequent and efficiently produce water as well as methanol. Deuteration  
reactions are also fast but due to its lower gas phase abundance in comparison
to the atomic hydrogen, production of deuterated species are less efficient than the hydrogenated species.
In case of Set 2, experimentally obtained energy barriers for olivine grain is used. In this case, atomic
hydrogen moves much slower than Set 1 energy barriers, resulting in lower production of 
hydrogenated species in comparison to the Set 1 energy barrier case. For Set 3, 
experimental values for the amorphous carbon grain were used and for Set 4, consideration
of \citet{lips04} is used. For both the cases, energy barriers   
against diffusion of hydrogen atom is even higher than Set 2, which results in the
insignificant production of methanol. Thus, most of the accreted gas phase CO 
remain unutilized. Due to the higher desorption energy of $\mathrm{CO}$, it
is trapped on the grain and cover significant percentage of grain mantles in 
case of Set 3 and Set 4. $\mathrm{O_3}$ occupies a major portion of the grain mantle
in case of set 3 and set 4 energy barriers. 

\subsection{Effects of irradiation}

Effects of interstellar radiation field on grain mantles were already discussed by
\citet{das11}. They pointed out that the choice of number of affected 
layers could severely affect the morphology of interstellar grain mantles. 
Interstellar photons could penetrate deep inside the interstellar ice. 
So, it is essential to have an idea about the stopping power as well as its range
into various ices. In order to see irradiation effects, we study the
properties of $0.4$ MeV projectiles for various target materials. These calculations are 
performed by using Ziegler’s SRIM program \citep{zieg03}. 
We use $\mathrm{H}^+$, $\mathrm{Fe}^+$, $\mathrm{S^+}$ and $\mathrm{C^+}$ as the projectiles. 
For the target ice, we consider
various types of ices shown in Table 6. It is clear from Table 4, that the grain mantle 
would be mainly covered by $\mathrm{H_2O,\ CH_3OH}$, their isotopes and $\mathrm{CO_2}$.
For one case of Table 6, we choose, mixed ice as the target ice. Following \citet{das11,maju12},
here, we assume that mixed ice is made up of $70\%$ water, $20\%$ methanol and 
$10\%$ carbon-di oxide. In Table 6, we have shown the stopping powers (in MeV cm$^2$ gm$^{-1}$) and 
projected ranges (in $\mu$m) of these projectiles for various target ices and projectiles. 
According to theory of transport of ions in matter (TRIM) prescribed by 
\citet{zieg03}, SRIM/TRIM program always makes a guess for the target density by 
considering a mixture of elemental target densities weighted by their relative stoichiometry. 
\citet{huds04} choose the target density as $1$ gm/cm$^3$ for water ice. 
Since here we compare between various targets under same projectile energy, 
we keep this density also the same for all the targets.
Stopping of ions are different from the weighted stopping of ions in the elemental matter which makes 
up the compounds \citep{zieg98}. For some target materials this correction is automatically included in 
the program developed by \citet{zieg98}.
These corrections are $0.94$, $0.9470706$ and $1.0$ for water, methanol and carbon-di-oxide respectively.
Since for the mixed ice no correction was included in the program, we consider it to be $1.0$ in 
our calculation. It is clear from Table 6 that
$\mathrm{Fe}^+$ has more stopping power than $\mathrm{H}^+$ but its projected range is lower. However,
both the ranges are longer than a typical grain size ($0.1$ $\mu$m). So, these projectiles
would easily penetrate deep into the ice layers for all of our cases.
 
\begin{table*}
\centering{ 
\scriptsize
\caption{Properties of $0.4$ MeV projectile and target material with a target density of $1.0$ gm/cm$^3$.}
\begin{tabular}{|c|c|c|c|}
\hline
{\bf Ice}&{\bf Projectile}&{\bf Stopping Power}&{\bf Range}\\
&&{\bf (MeV cm$^2$ gm$^{-1}$)}&{\bf ($\mu$m)}\\
\hline
\hline
&$\mathrm{H^+}$&425.8/403.3/383.0&6.68/7.06/7.43\\
$\mathrm{\bf H_2O/HDO/D_2O}$&$\mathrm{Fe^+}$&1905/1804/1714&0.6253/0.6731/0.7230\\
&$\mathrm{S^+}$&2819/2669/2535&0.9569/1.03/1.11\\
&$\mathrm{C^+}$&2875/2722/2585&1.86/1.98/2.11\\
\hline
&$\mathrm{H^+}$&461.1/447.1/433.8&6.09/6.28/6.48\\
$\mathrm{\bf CH_3OH/singly \ deuterated/doubly \ deuterated}$&$\mathrm{Fe^+}$&2591/2512/2438&0.5648/0.5881/0.6118\\
&$\mathrm{S^+}$&3678/3566/3461&0.8117/0.8451/0.8791\\
&$\mathrm{C^+}$&3427/3322/3224&1.52/1.58/1.63\\
\hline
&$\mathrm{H^+}$&364.8&8.28\\
$\mathrm{\bf CO_2}$&$\mathrm{Fe^+}$&1890&0.7225\\
&$\mathrm{S^+}$&2712&1.07\\
&$\mathrm{C^+}$&2679&2.01\\
\hline
&$\mathrm{H^+}$&445.4&6.42\\
{\bf Mixed ice}&$\mathrm{Fe^+}$&3425&0.6151\\
&$\mathrm{S^+}$&3186&0.9075\\
&$\mathrm{C^+}$&3138&1.71\\
\hline
\end{tabular}}
\end{table*}

\section{Conclusions}

We carried out Monte Carlo simulation to study chemical composition of interstellar grain mantles
under various physical circumstances. We summarize our conclusions bellow:

\noindent $\bullet$ Despite low elemental abundance of atomic deuterium in interstellar space, 
we find that it could significantly alter the chemical composition of grain mantles. 

\noindent $\bullet$ Binding energies control surface composition of grains. We
considered various types of binding energies to study their effects. It is noted that production
of surface species is favourable when we use the `Set 1' energy barriers.

\noindent $\bullet$ Interstellar radiation field strongly affects chemical composition of
interstellar grain mantles. Our results clearly showed that in diffuse clouds, 
hardly a few layers of surface species are formed. We also showed that
in translucent cloud region, surface coverage of interstellar grain mantles is mainly 
dominated by photo dissociation products ($\mathrm{C,CH_3,CH_2D,OH}$ and OD).
Around intermediate dense cloud region, water and methanol with their deuterated forms 
cover major portion of grain mantles. In deep interior (higher density region),
formation of oxygenated species ($\mathrm{CO_2,O_2,O_3}$) increases and significant part of the
grain mantles would be covered by $\mathrm{CO}$ molecules as well. 

\noindent $\bullet$ A comparison between our calculated deuterium fractionation ratio and
observed fractionation ratio was made. We obtained a very high degree of deuterium 
fractionation for water and methanol. Observational evidences also supports this
high fractionation feature of methanol but surprisingly, very low fractionation
ratio were observed for water molecules. This is in the contradiction with our results. 
Possible cause of this discrepancy was already discussed by \citet{pari03}. However, a satisfactory
explanation is yet to be found.

\noindent $\bullet$ We used $\mathrm{H^+,\ Fe^+,\ S^+}$ and C$^+$ ions to study 
the stopping power and projected range on various types of ice targets 
such as normal water, deuterated water, normal methanol, deuterated methanol and mixed ice. We find that 
they do penetrate in the mantles and are likely to heat and evaporate them, apart from causing 
new ice and ice phase reactions. These will be studied in future.

\section{Acknowledgments}
AD, DS \& SKC are grateful to ISRO respond (Grant No. ISRO/RES/2/372/11-12), AD \& SKC
want to thank DST (Grant No. SB/S2/HEP-021/2013) for financial support. 
LM thanks MOES and ERC starting grant (3DICE, grant agreement 336474) for funding during this work.

\clearpage

\begin {figure}
\vskip 2cm
\hskip -2cm
\centering{
\includegraphics[height=13cm,width=20cm]{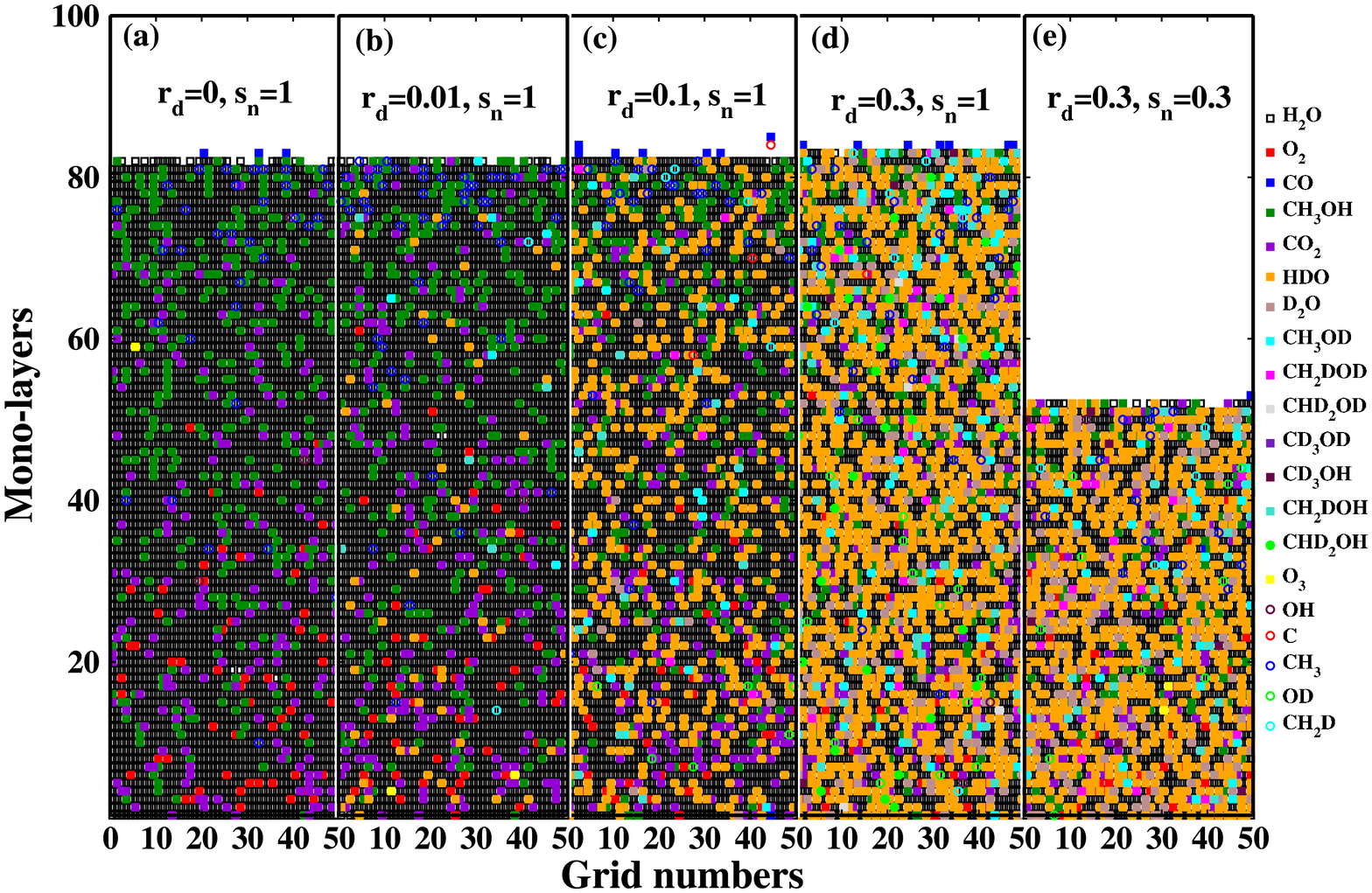}}
\caption{{\bf(a-e)} A cross-sectional view of the composition of grain mantle 
with the (a) absence and (b-e) presence of deuterium.}
\label{fig-2}
\end {figure}
\clearpage

\begin {figure}
\vskip 2cm
\centering{
\includegraphics[height=10cm,width=10cm]{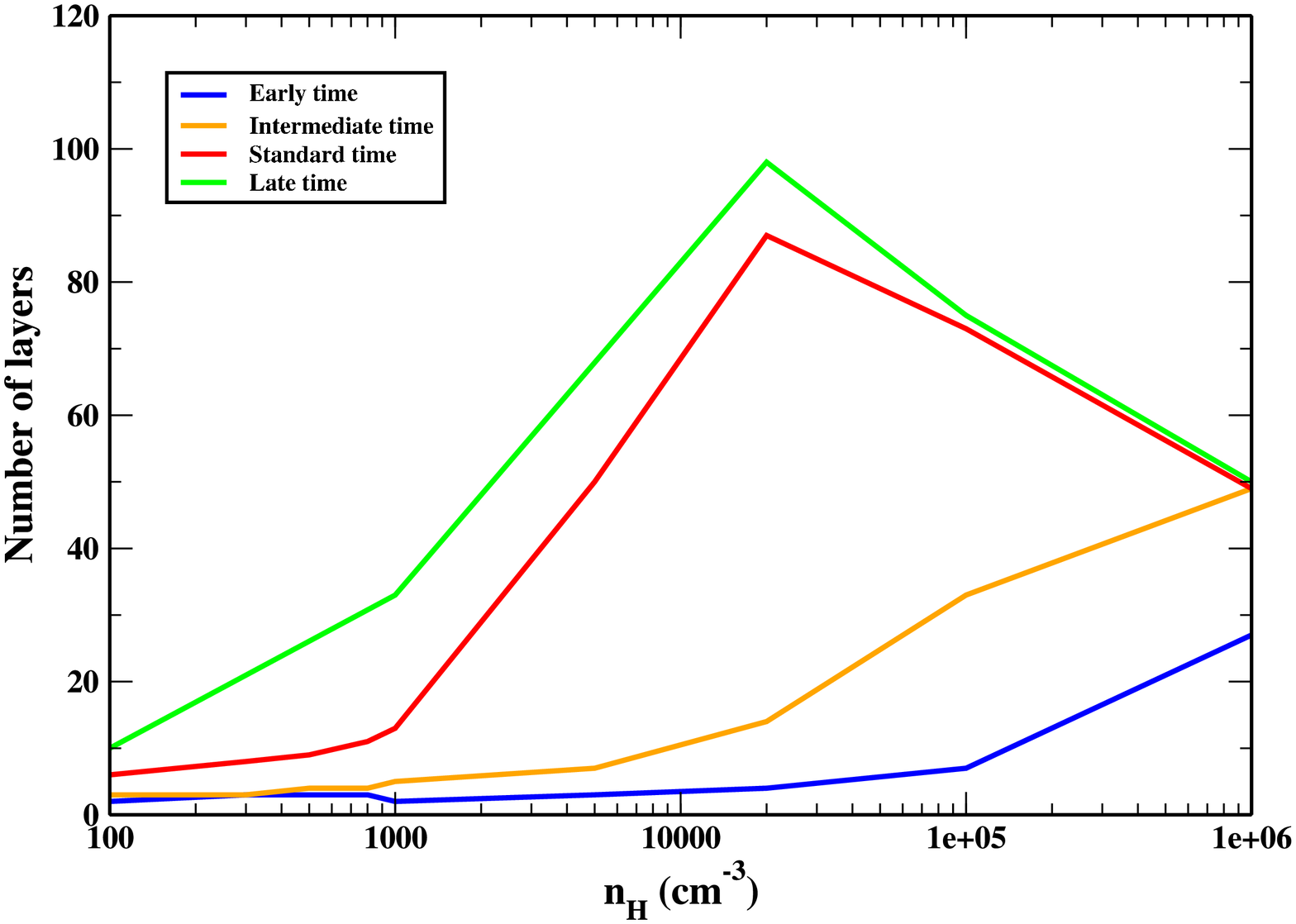}
}
\caption{Formation of numbers of layers in various stages of chemcal evolution.}
\label{fig-3}
\end {figure}
\clearpage

\begin {figure}
\vskip 2cm
\hskip 0.5cm
\centering{
\includegraphics[height=14cm,width=14cm]{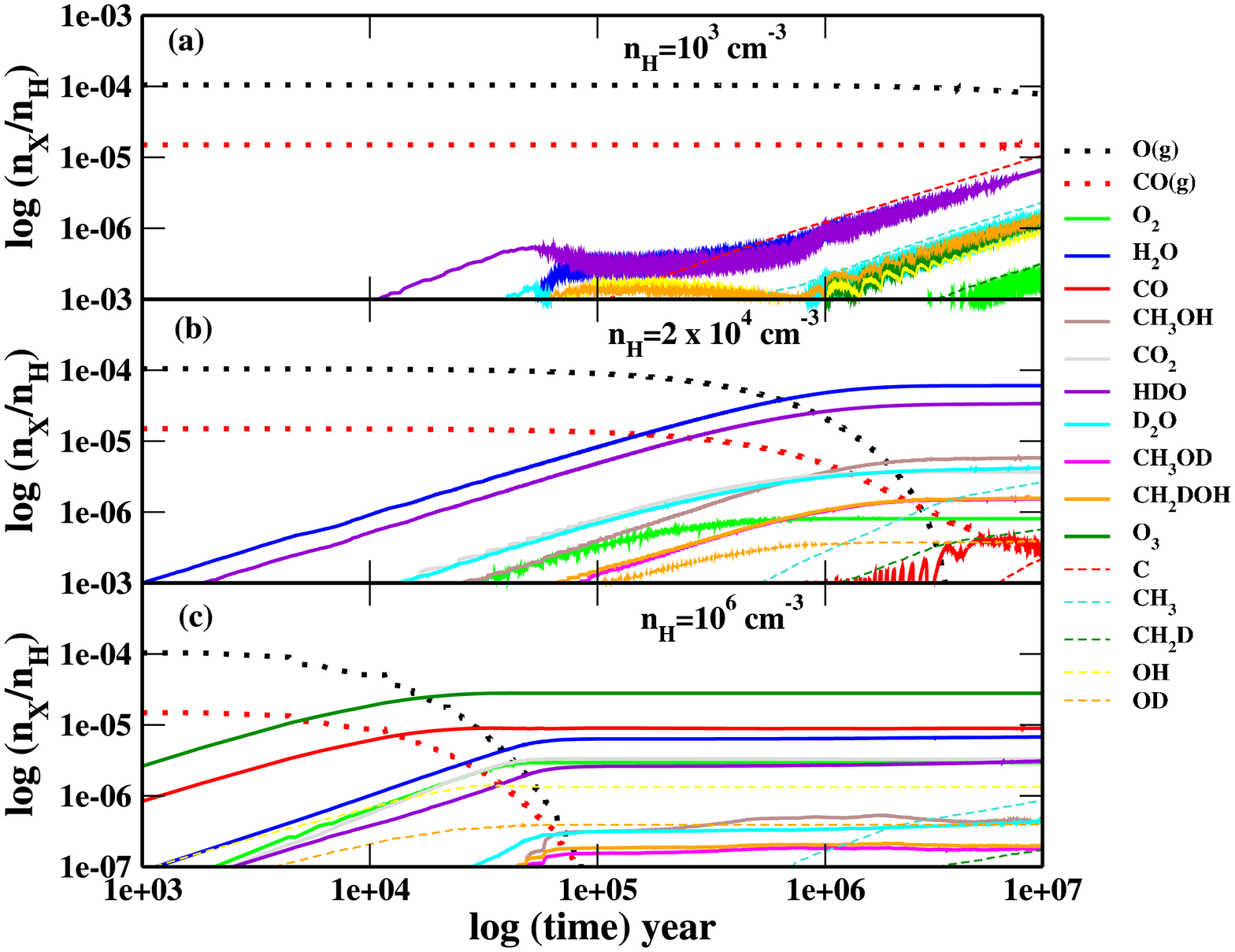}
}
\caption{{\bf(a-c)} Chemical evolution of important gas/ice phase species for 
(a)$n_H=10^4$ cm$^-3$, (b)$n_H=10^5$ cm$^{-3}$ and (c)$n_H=10^6$ cm$^{-3}$.}
\label{fig-4}
\end {figure}
\clearpage

\begin{figure}
\vskip 1cm
\centering{
\includegraphics[height=14cm,width=20cm]{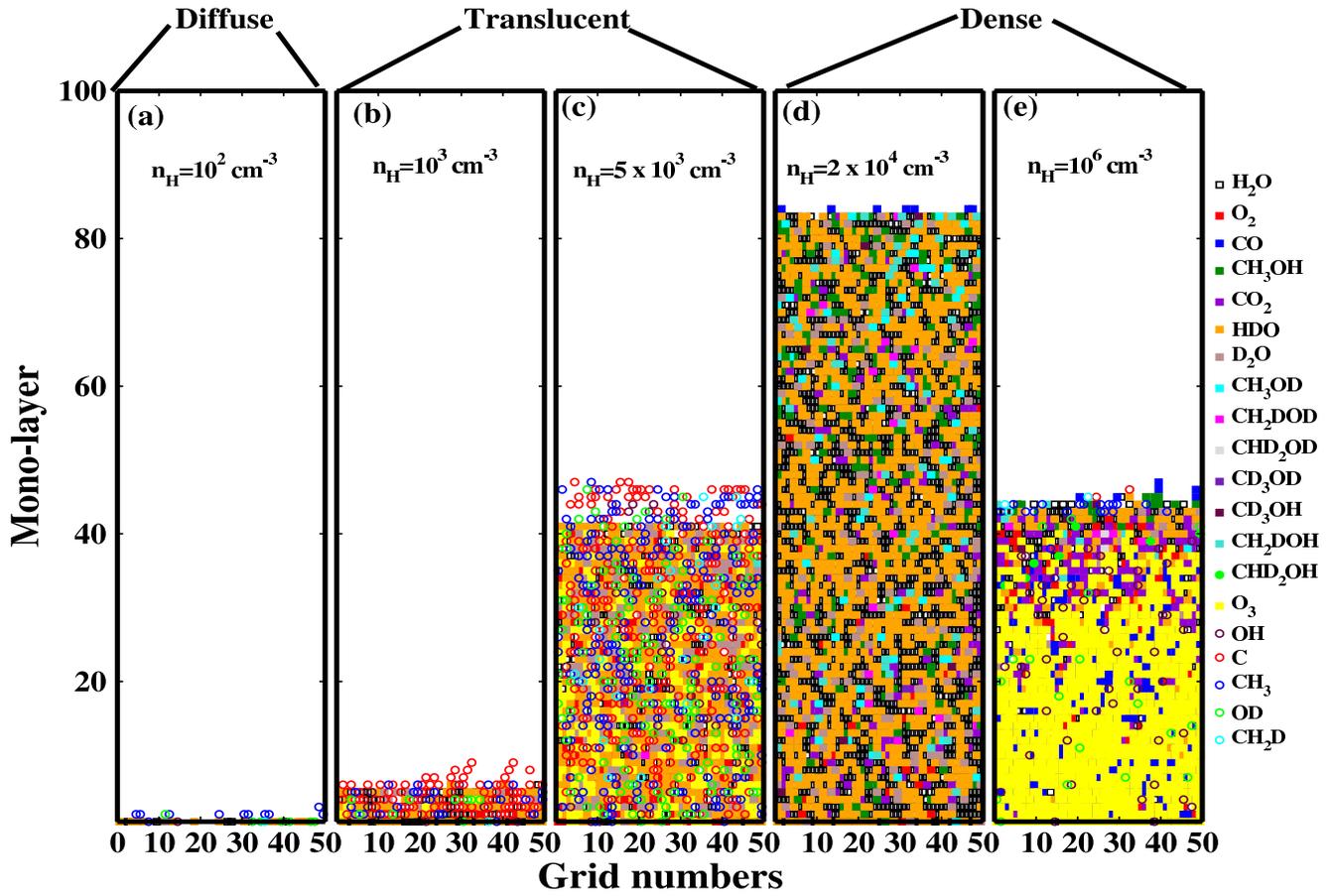}
\caption{{\bf (a-e)} Composition of grain mantle for various types of clouds.}
\label{fig-5}}
\end {figure}
\clearpage

\begin{figure}
\vskip 2cm
\centering{
\includegraphics[height=10cm,width=10cm]{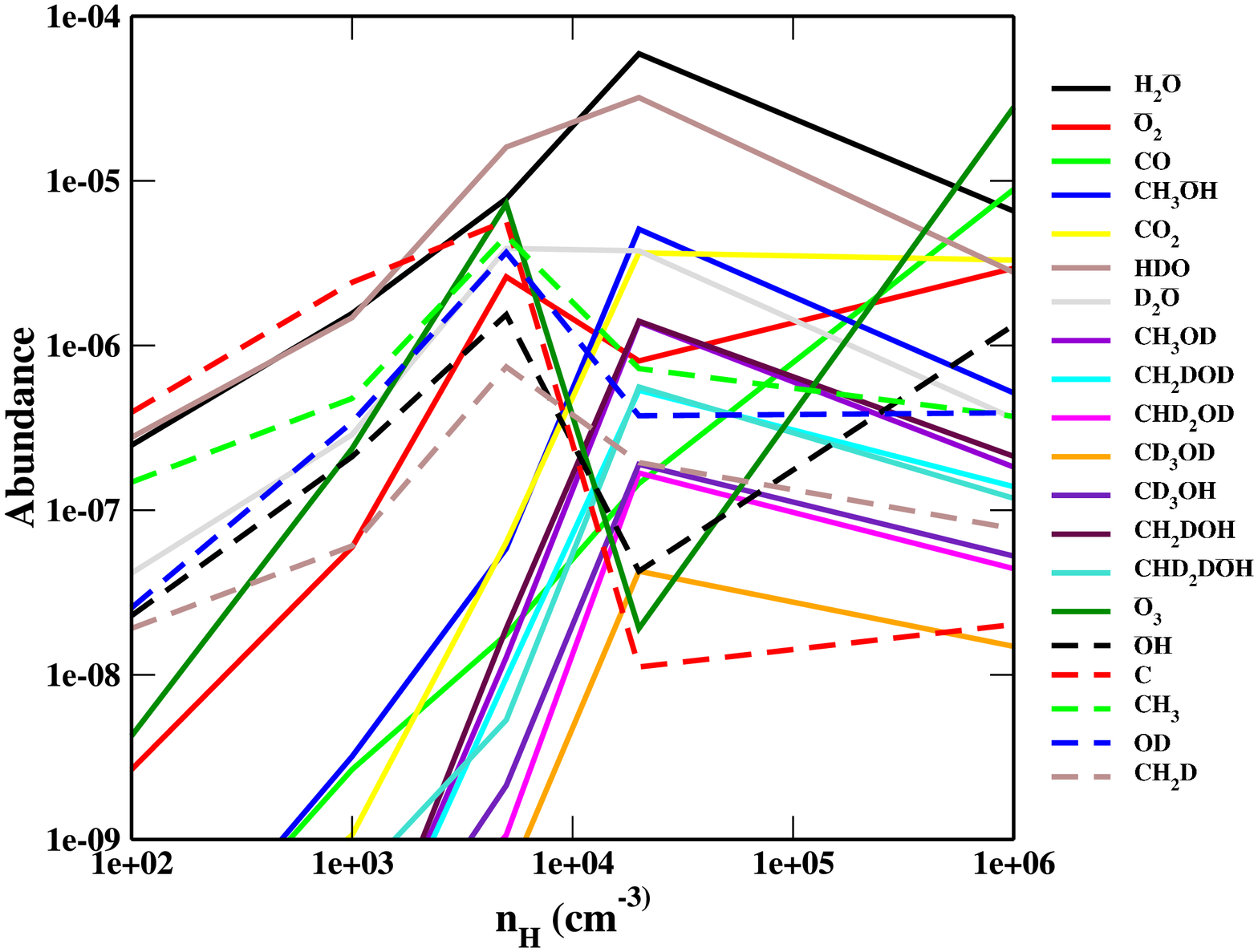}
\caption{ Abundances of some important surface species around various clouds.}
\label{fig-6}}
\end {figure}
\clearpage

\begin{figure}
\vskip 2cm
\centering{
\includegraphics[height=13cm,width=18cm]{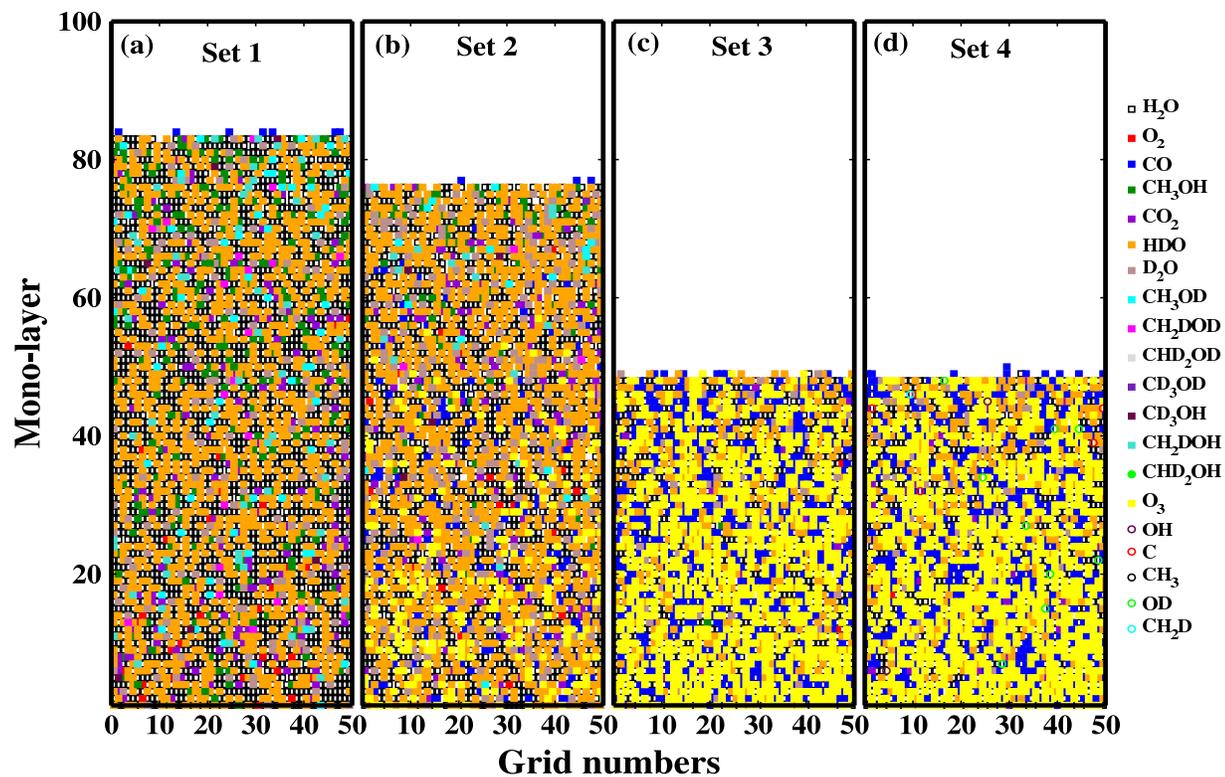}
\caption{{\bf(a-d)} Composition of grain mantle for various sets of binding energies.}
\label{fig-7}}
\end {figure}
\end{document}